\documentclass[fleqn,twoside]{article}
\usepackage[headings]{espcrc2}

\readRCS
$Id: qcd04.tex,v 1.1 2004/07/11 16:26:08 schorner Exp $
\usepackage{graphicx}
\usepackage[figuresright]{rotating}

\newcommand{\AmS}{{\protect\the\textfont2
  A\kern-.1667em\lower.5ex\hbox{M}\kern-.125emS}}

\title{Jets and the hadronic final state at HERA}

\author{T. Sch\"orner-Sadenius\address[UHH]{Universit\"at Hamburg, IExpPh,  
	Luruper Chaussee 149, 22761 Hamburg, Germany} 
  on behalf of the H1 and ZEUS collaborations}
       
\runtitle{Jets and the hadronic final state at HERA}
\runauthor{T. Sch\"orner-Sadenius}

\begin{document}

\begin{abstract}
Recent results on jets and the hadronic final state from the HERA
collaborations H1 and ZEUS are reviewed.
\vspace{1pc}
\end{abstract}

\maketitle

\section{INTRODUCTION}

Until the year 2000, the HERA experiments H1 and ZEUS have collected
integrated luminosities of about 130~pb$^{-1}$. A large number of
interesting results on jets and the hadronic final state from this `HERA 1'
data taking period have already been published, and many more
analyses are still ongoing. 

In 2001/2002 the HERA machine and also the experiments have undergone major
changes with the aim of increasing the delivered luminosity by a factor of
five. Due to technical problems, efficient data taking could not start before
the end of 2003. Therefore, new results from the `HERA 2' data taking period
are only now starting to come out. 

In the recent years the analysis work on the HERA 1 data has concentrated on
making optimal use of the high statistics and either decreasing statistical
(and also systematic) errors or performing more and more differential
measurements. 
In addition, increased statistics allows for the discovery or first
measurement of new or exotic phenomena. 

This contribution will highlight new results concerning jets and the
hadronic final state at HERA which were published or made preliminary since the
EPS03 conference in Aachen. 

\begin{figure}[t]
\begin{center}
\includegraphics*[width=0.45\textwidth]{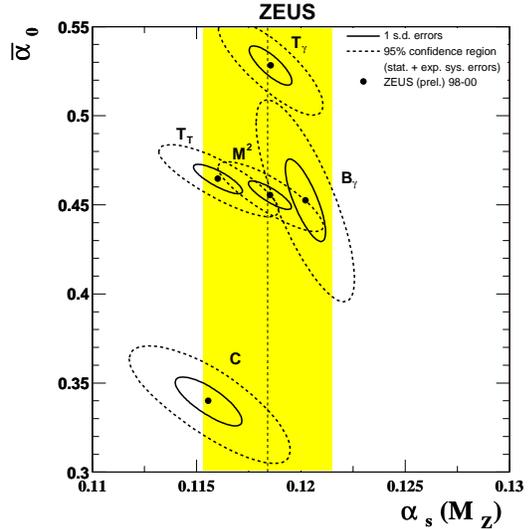}
\vspace*{-0.75cm}
\caption{\label{eventshapes} ZEUS event shape variables: Simultaneous fit of
  $\alpha_S$ and the PC parameter $\overline{\alpha_0}$ to the mean of event shape
  variables using NLO+PC predictions.}
\end{center}
\end{figure}

\section{\label{hadro}EVENT SHAPE VARIABLES}

Event shape variables like the thrust $T$, broadening $B$ or the $C$ parameter
measure 
aspects of the topology of an event's hadronic final state. These variables
are inclusive and infrared safe so that hadronisation corrections can be
estimated using power correction (PC) models. 

ZEUS has performed measurements of event shape variables in the Breit frame in
the range 80~$< Q^2 <$~20480~${\rm GeV^2}$~\cite{refshape}. 
The choice of the Breit frame
assures that the current region, i.e. the direction of flight of the scattered parton,
resembles a single hemisphere in $e^+e^-$ collisions, thus making the HERA
results comparable to results from other colliders, especially LEP. 

The PC model suggests that non-perturbative corrections to
the 
predictions of observables can be parametrised with a universal parameter 
$\overline{\alpha_0}$ according to
$\langle F \rangle = \langle F \rangle_{NLO} + \langle F \rangle_{pow}(\alpha_S,\overline{\alpha_0})$.
So if the next-to-leading (NLO) prediction $\langle F \rangle_{NLO}$ exists, one can fit for
$\alpha_S$ and for the 
universal parameter $\overline{\alpha_0}$.

In addition to the power corrections, next-to-leading log (NLL) resummations
can help to improve the description of differential event shape
distributions. 
The use of NLL sums requires a matching with the NLO prediction in
order to avoid double-counting of terms. 

Fitting the means of the event shape variables with NLO+PC predictions 
one obtains the results for $\alpha_S$ and $\overline{\alpha_0}$ shown in 
Figure~\ref{eventshapes}. 
The resulting $\alpha_S$ values are consistent with the current world average,
$\alpha_S =$~0.118. The data suggest $\overline{\alpha_0} =$~0.5  
consistently for all event shape variables except for the $C$ parameter, the
fit of which is extremely sensitive to the fit range in the ZEUS analysis (in
contrast to H1 measurements). This result for $\overline{\alpha_0}$ is roughly
compatible with the results from other experiments. 

\section{\label{prompt}PROMPT PHOTON PRODUCTION}

The ZEUS collaboration has for the first time measured prompt photon
production in DIS in the full HERA 1 data sample of about 
120~${\rm pb^{-1}}$~\cite{zeusprompt}. Prompt photons are required to be
well-isolated and are identified using
calorimeter cluster shape information; neutral  particle
backgrounds are subtracted using Monte Carlo (MC) predictions. 

\begin{figure}[h]
\begin{center}
\includegraphics*[width=0.35\textwidth]{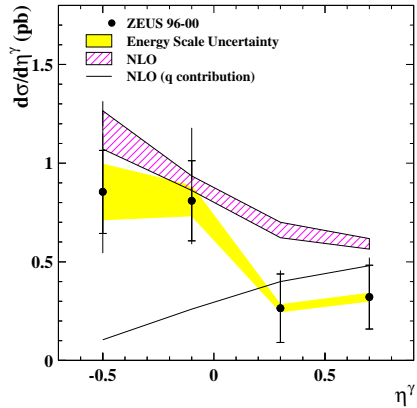}
\vspace*{-0.75cm}
\caption{\label{prompt1}ZEUS prompt photon plus jet sample: photon pseudo-rapidity
  distribution. The data, which are shown together with the energy scale
  uncertainty band, are compared to the NLO calculations for which the
  renormalisation scale uncertainty is also indicated. Also shown is the
  contribution to the cross-section which is due to prompt photon emission
  from the quark line.}
\end{center}
\end{figure}

Two analyses were applied to the data:
In the first, inclusive prompt photon production is studied, and the data are
compared to the leading-order (LO) MC models HERWIG and PYTHIA; none of the models is able to
describe all features of the data, and both are far away from the data in
terms of normalisation.

In the second analysis, a jet in the central rapidity region was required in
addition to the isolated photon. Jets were reconstructed using a cone
algorithm in the laboratory frame; their transverse energy had to be in excess
of 6~GeV, and their pseudo-rapidity between -1.5 and 1.8. 
The measurements, which yield a total
cross-section of 0.86~pb for prompt photon plus jets, are again compared to
HERWIG 
and PYTHIA, but there exists also an NLO QCD calculation for this signature. 
PYTHIA and HERWIG manage to describe the shapes of the photon and jet
transverse energy distributions, but they fail in describing both
pseudo-rapidity distributions. In addition, their normalisation is again of by
factors of about 2 (4) for PYTHIA (HERWIG). 

The NLO QCD calculation for the prompt photon plus jet analysis predicts a
cross-section of about 1.33~pb, which is supposed to be lowered by
hadronisation corrections by about 30-40~$\%$. The normalisation of the
calculation is therefore compatible with the data. 
As an example, Figure~\ref{prompt1} shows 
the prompt photon pseudo-rapidity distribution. The data are in agreement with
the NLO predictions; however, the statistical uncertainty of the measurement is
very large. The same is true for the jet transverse energy
distribution. However, the calculations
predict a too large cross-section at low photon transverse energies and at high
(forward) jet pseudo-rapidities.

\begin{figure}[h]
\begin{center}
\includegraphics*[width=0.4\textwidth]{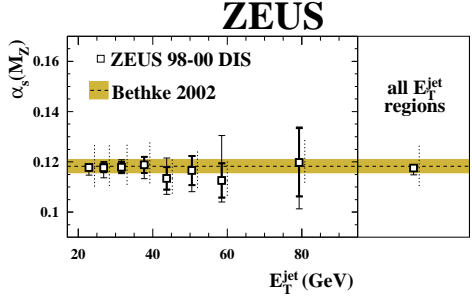}
\vspace*{-0.75cm}
\caption{\label{shape2}ZEUS $\alpha_S$ measurements based on the mean 
integrated jet shape $\langle \psi(r=0.5) \rangle$
in DIS events as a function of the jet transverse
energy. The data are shown together with the combined $\alpha_S$ value and the
world average.}
\end{center}
\end{figure}

\section{\label{precise}NEWS FROM JET PHYSICS }

\subsection{Determinations of $\alpha_s$ from jet shapes}

The ZEUS collaboration has measured the substructure dependence of jet
cross-sections in photoproduction and DIS with the aim of extracting the
strong coupling parameter $\alpha_S$~\cite{zeusshapes}. 
ZEUS used mainly the mean integrated
jet shape $\langle \psi(r) \rangle$ which is defined as: 

\begin{equation}
\langle \psi(r) \rangle = \frac{1}{N_{jets}}\Sigma_{jets}\frac{E_T(r)}{E_T^{jet}}.
\end{equation}  
Here, $N_{jets}$ is the total number of jets in the sample, $E_T(r)$ is the
transverse energy within a cone of radius $r$ around the jet axis, and
$E_T^{jet}$ is the transverse energy of the jet. The jet shape was measured,
for example, as a function of the jet transverse energy in DIS events. For
each data point, 
an $\alpha_S$ value is determined (Figure~\ref{shape2}); 
the resulting values are evolved to the
mass of the $Z$, and the different values are combined into one
$\alpha_S(M_Z)$ value:  
$\alpha_S = 0.1176\pm^{0.0091}_{0.0072}$, where only the dominating
theoretical uncertainty is given.

\subsection{Summary on $\alpha_S$}

\begin{figure}[h]
\begin{center}
\includegraphics*[width=0.4\textwidth]{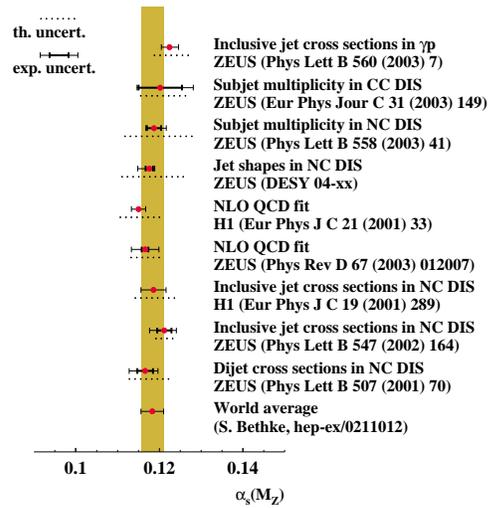}
\vspace*{-0.75cm}
\caption{\label{alpscollection}A summary of $\alpha_S$ measurements from H1
  and ZEUS jet analyses.}
\end{center}
\end{figure}

Figure~\ref{alpscollection} gives an overview of various $\alpha_S$
measurements performed by the HERA collaborations in jet events. The
measurements are in perfect agreement with the world average; the already very
good precision of the measurements, which has been increased over the past few
years, will be even more increased once the HERA 2 data are collected and
analysed. 

\subsection{Jets in global QCD fits}

Jet data from the HERA experiments can be used for more than just
determinations of the strong coupling. Recently, the ZEUS collaboration has
started to include jet data into their global QCD fits ~\cite{zeusglobal},
thus supplementing the 
usual $F_2$ data. The aim of using jet data is to improve the precision of the
PDF measurements, especially the gluon, at high values of $x$.

\begin{figure}[h]
\begin{center}
\includegraphics*[width=0.45\textwidth]{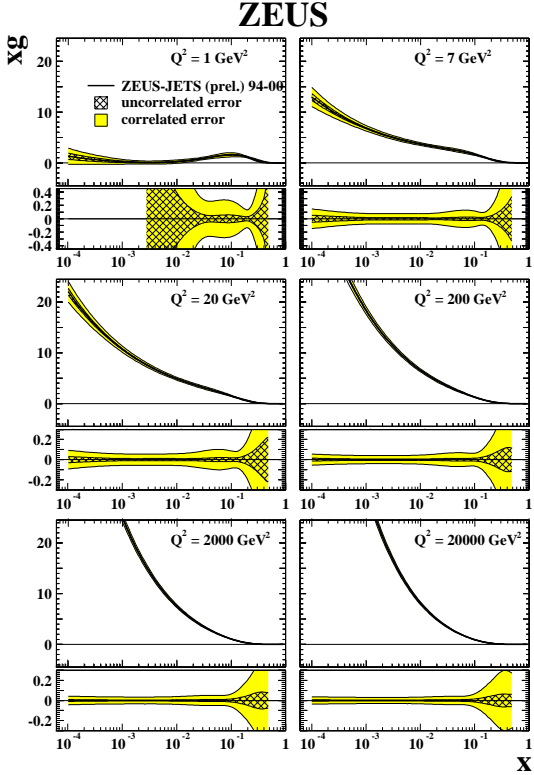}
\vspace*{-1.25cm}
\caption{\label{zeusjets2}ZEUS global QCD fit. The gluon
  density as a function of $x$ is shown for various $Q^2$ bins. Presented are
  the 
  results of the ZEUS fit including jet data.}
\end{center}
\end{figure}

The use of jet data in the fits, however, requires a fast evaluation of the NLO
predictions for the jet cross-sections. During the
minimisation procedure, this prediction has to be performed many thousand
times. The usual typical 8 hours for sufficient statistics clearly is not
acceptable here. The way around this 
problem is based on the following method. The jet cross-section can be written
as the convolution of the hard partonic cross-section $\hat{\sigma}$ with the
PDF $f_a$:
\begin{equation}
\sigma = \Sigma_a \Sigma_{n=1}^{\infty}\alpha_S^n \int d\eta
f_a(\eta,\mu_F) \cdot \hat{\sigma}_n(\eta,\mu_F)
\end{equation} 
Here, the first sum runs over all parton flavours. 
Defining a sufficiently finely binned grid in $(\eta,\mu_F)$ and assuming
that the PDF $f_a(\eta,\mu_F)$ is flat in every bin separately, one can write
the cross-section in one bin of the $(\eta,\mu_F)$ grid as:
\begin{equation}
\label{fl2}
\sigma(\eta,\mu_F) \sim \Sigma_a f_a(\eta,\mu_F) \cdot \Sigma_{n=0}^{\infty}
\alpha_S^n \int d\eta \hat{\sigma}_n(\eta,\mu_F)
\end{equation}
where the integration is now only over the corresponding bin in $\eta$. The
summation of all pieces $\sigma(\eta,\mu_F)$ leads again to the total
cross-section. 
The benefit of this method is that the coefficients (the integrals in
Equation~\ref{fl2}) have to be evaluated only once since they are independent
of the PDFs; they can thus be stored in a grid and be convoluted with any PDF
provided from the outside, for example from the fitting program. The
convolution then takes only parts of a second, which allows for the use of
this cross-section evaluation in fitting programs. 

The effect of the use of jet data can be viewed in
Figure~\ref{zeusjets2} which shows the
results for the gluon density as achieved by the ZEUS fit using
inclusive ($F_2$) and jet data. The data are shown as function of $x$ in different bins
of $Q^2$. A clear decrease of the size of the uncertainty for the low-$Q^2$ region
is found; the effect is less striking for higher $Q^2$ values.  

For the future, ZEUS aims at including further jet data from photoproduction,
heavy flavour production and also, if possible, from proton-antiproton
collisions into the QCD fit. 

\subsection{\label{problems}FROM PHOTOPRODUCTION TO DIS}

The transition region from photoproduction to DIS has since long been
problematic in the theoretical description. The HERA collaborations have
published new results on this
topic~\cite{transition1,transition2,transition3}. 

\begin{figure}[t]
\begin{center}
\includegraphics*[width=0.4\textwidth]{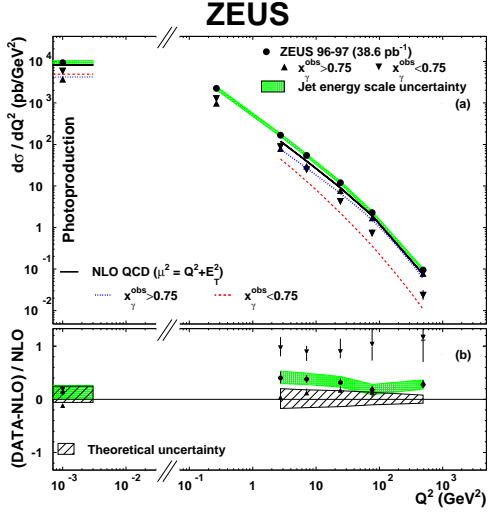}
\vspace*{-0.75cm}
\caption{\label{resolvedzeus}ZEUS dijet cross-section as a function of
  $Q^2$. Separately shown are the direct- and resolved-enriched samples and
  their description by the NLO QCD calculation. }
\end{center}
\end{figure}

\begin{figure}[h]
\begin{center}
\includegraphics*[width=0.45\textwidth]{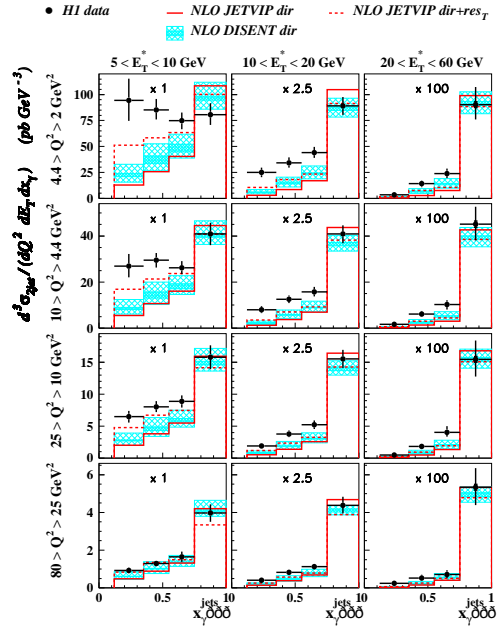}
\vspace*{-0.75cm}
\caption{\label{kamil}H1 triple-differential dijet cross-section.}
\end{center}
\end{figure}

H1 has analysed dijet cross-sections in the kinematic
region 5~$< Q^2 <$~100~${\rm GeV^2}$~\cite{transition1}. The data, taken in
the years 
1996 and 1997, are compared to direct NLO calculations. 
Overall good agreement between
data and predictions is observed. However, comparing the
data to NLO predictions for dijet events with small azimuthal separation
between the two hardest jets shows a clear excess of the data over the
prediction. This 
discrepancy can be cured partly using a NLO three-jet calculation; however
discrepancies remain at low $Q^2$ and $x$. 

\begin{figure}[t]
\begin{center}
\includegraphics*[width=0.3\textwidth]{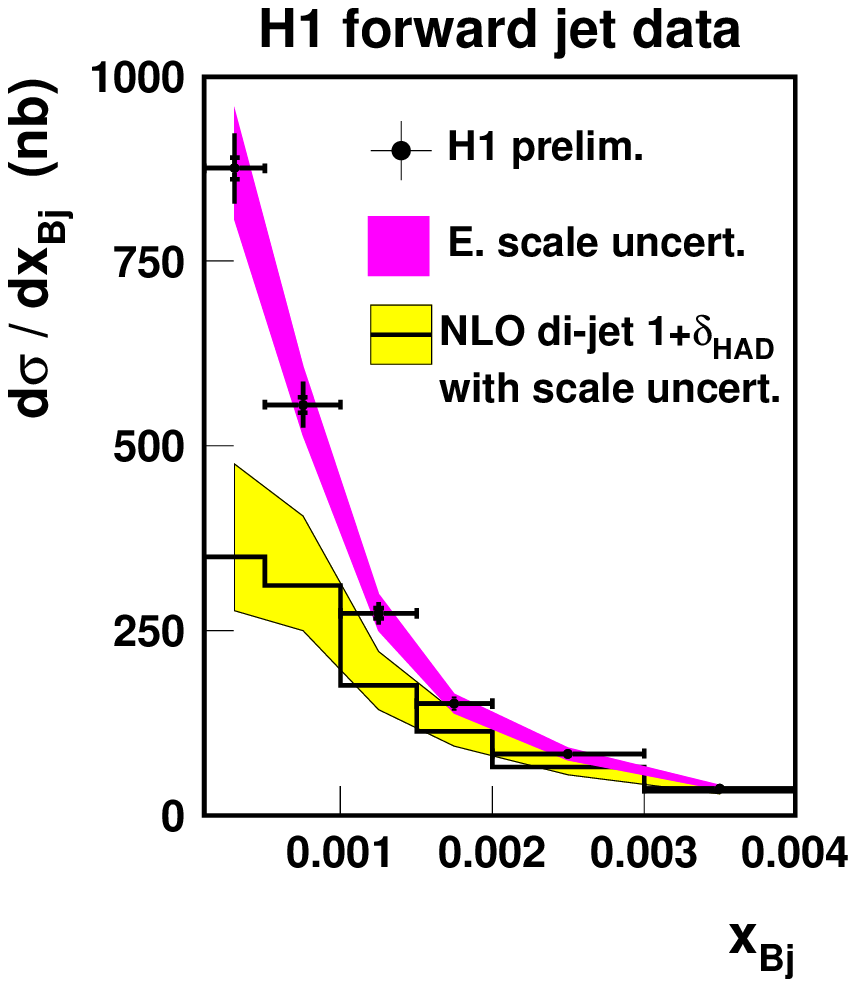}
\vspace*{-0.75cm}
\caption{\label{h1forward}H1 forward jet measurement as a function of $x$. The
data are compared to (direct) NLO QCD calculations. Also indicated are the
energy scale and renormalisation scale uncertainties.}
\end{center}
\end{figure}

Figure~\ref{resolvedzeus} shows the ZEUS dijet cross-section as a function of
$Q^2$ over a wide kinematic range from 0~${\rm GeV^2}$ to 
2000~${\rm GeV^2}$~\cite{transition2}. 
The data are separated into regions of direct- and  
resolved-enriched events by means of the observable $x_{\gamma}$ which
quantifies the fraction of the photon's energy that entered the hard
scattering. Shown is also the NLO QCD prediction. For the DIS regime, the NLO
calculation, which 
does not include resolved contributions, clearly fails to describe the
low-$x_{\gamma}$ component (independent of $Q^2$) and also the low-$Q^2$
region.  
However, the prediction is very sensitive to the 
renormalisation scale choice, and scale choices can be made such that the data
are compatible with the predictions. For the photoproduction region, which
also is dominated by low $x_{\gamma}$ values, the (direct+resolved) NLO
calculation is able to describe the data. 

H1 has measured the triple-differential dijet cross-section 
$d^3\sigma/d\overline{E}_TdQ^2dx_{\gamma}$~\cite{transition3} 
($\overline{E}_T$ is the mean transverse energy of the two jets). 
Also here the conclusion is 
that at low $Q^2$ and also at low values of 
$\overline{E}_T$ NLO QCD calculations fail to describe the data at low values
of $x_{\gamma}$. This is the case even if the calculations include the
contributions from the resolved photon (Figure~\ref{kamil}). 
In the same analysis, H1 could also demonstrate the importance of the initial
and final state parton showers which help considerably in describing the data
with LO MC models.   

\subsection{\label{problems2}PARTON DYNAMICS AT LOW $x$}

Another problematic issue in jet physics is the question of parton dynamics at
low $x$. Earlier studies of so-called forward jets~\cite{forwardjetsold} were
designed to hunt for the break-down of DGLAP evolution and the onset of BFKL
signatures. However, the results were never really conclusive, 
mainly because no NNLO DGLAP calculation and no BFKL prediction
usable for experimentalists were (and are) available.  Also the influence of
resolved contributions to the predictions was unclear.

\begin{figure}[h]
\begin{center}
\includegraphics*[width=0.45\textwidth]{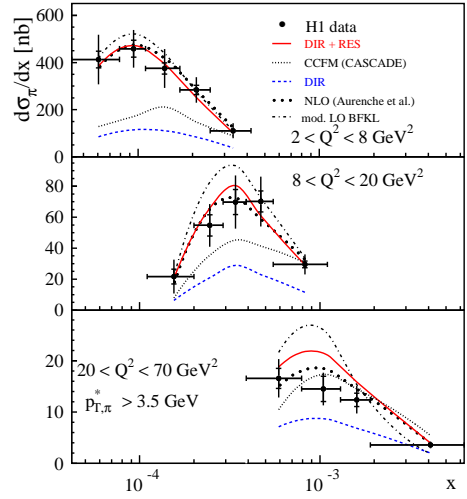}
\vspace*{-0.75cm}
\caption{\label{forwardpi0}H1 forward pion measurement as a function of
  $x$. The data are compared to various QCD calculations and models.}
\end{center}
\end{figure}

There are new results by the H1 collaboration on forward jet and $\pi^0$
production in low $Q^2$ DIS~\cite{h1forwardnew1,h1forwardnew2}. 
The forward jet analysis, which was
performed in the kinematic region 5~$< Q^2 <$~85~${\rm GeV^2}$, studied the
triple-differential cross-section $d^3\sigma/d\overline{p}_T^2dQ^2dx$ and
compared the data to NLO QCD calculations and various MC
models~\cite{h1forwardnew1}. One result is shown in Figure~\ref{h1forward},
where the data are presented as a function of $x$ 
and are compared to the NLO QCD calculation which clearly fails to describe
the data at low values of $x$. MC models implementing resolved photon
contributions and  models using the the colour-dipole model come close to the
data, although a discrepancy between data and predictions remains at lowest
values of $x$. A more differential study shows that the discrepancies are
largest for low transverse energies and low $Q^2$.  

For the case of forward $\pi^0$ production~\cite{h1forwardnew2}, 
most of the available predictions 
do a sufficiently good job. 
Figure~\ref{forwardpi0} shows the forward $\pi^0$ cross-section as a
function of $x$ in three bins of $Q^2$; the good description of the data by
most models (including the NLO calculation involving a convolution of matrix
elements and fragmentation functions) is prominent. 

\section{\label{pq}STRANGE PENTAQUARKS}

\begin{figure}[h]
\begin{center}
\includegraphics*[width=0.4\textwidth]{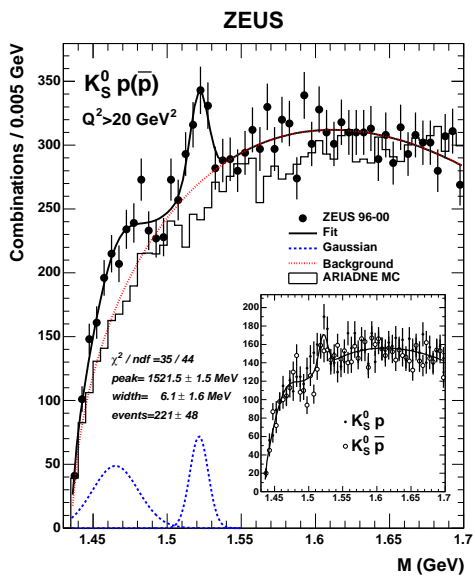}
\vspace*{-0.75cm}
\caption{\label{pqresult}The ZEUS $pK^0_S$ ($\overline{p}K^0_S$) invariant
  mass distribution with 
  the peak at about 1522~MeV. Also shown are the results of a fit with a
  background function and two gaussians, the background function and the
  simulation of the ARIADNE Monte Carlo.}
\end{center}
\end{figure}

The measurement of narrow baryonic resonances of positive strangeness in the
$K^+n$ and $K^0_Sp$ decay channels with a mass of about 1530~MeV and a very
narrow width has recently raised a lot of attention. 
The findings are consistent
with five-quark bound states, so-called pentaquarks,
$\Theta^+=uudd\overline{s}$. 

The ZEUS collaboration started a search for the strange (anti)pentaquark in
the $K^0_Sp(\overline{p})$ channel~\cite{pqzeus}, relying mainly on the
precise tracking information from the central drift chamber. The analysis
used all events with $Q^2 >$~1~${\rm GeV^2}$ of
the full 1996-2000 data statistics of 121~${\rm pb^{-1}}$, about
1.600.000 events. 

The $K^0_Sp(\overline{p})$ invariant mass was obtained by combining the
$K^0_S$ and (anti)proton candidates and fixing the kaon mass to the PDG
value. Figure~\ref{pqresult} shows the resulting pentaquark candidate mass
distribution 
together with the fit (a background function and two gaussians), the
background function and the prediction of the ARIADNE Monte Carlo. A clear
peak at 1522~MeV with a width of about 6~MeV is visible. The number of events
ascribed to
the signal is 221$\pm$48, of which 96$\pm$34 are in the antiproton
channel. Thus, not only could the pentaquark be confirmed, but also the first
observation of the antipentaquark with quark content
$\overline{u}\overline{u}\overline{d}\overline{d}s$ was made.  

The charmed pentaquark state that was observed by the H1 collaboration will be
discussed in another proceedings contribution~\cite{pqh1}.  

\section{\label{mix}OTHER SELECTED TOPICS}

\subsection{QCD instantons}

Instantons are non-perturbative fluctuations of the gauge fields in
non-abelian gauge theories. But although instantons are predicted by the
standard model, they have not been observed so
far. Especially QCD instantons, however, are supposed to have short-distance
implications at sufficiently low energies; they are, for example, supposed to
induce characteristic topologies in DIS $ep$ scattering events. 
Results on instanton searches have been
reported on by the H1 collaboration before~\cite{h1instanton}. Recently the
ZEUS collaboration has published the results of a search for instanton
events~\cite{zeusinstanton} performed in about 40~${\rm pb^{-1}}$ of data
collected in the years 1996/97. The data were selected according to 
$Q^2 >$~120~${\rm GeV^2}$. For the search for instanton events, several
discriminating techniques were applied which distinguish between standard NC
DIS events and instanton events using multi-dimensional cuts. Instanton events
were simulated using the QCDINS calculation which predicts an 
instanton cross-section of 8.9~pb. Under the assumption that all selected
data events are instanton events, an upper limit on the instanton production 
cross-section of $\sigma <$~26~pb was derived at 95~$\%$ C.L. 

\begin{figure}[h]
\begin{center}
\includegraphics*[width=0.35\textwidth]{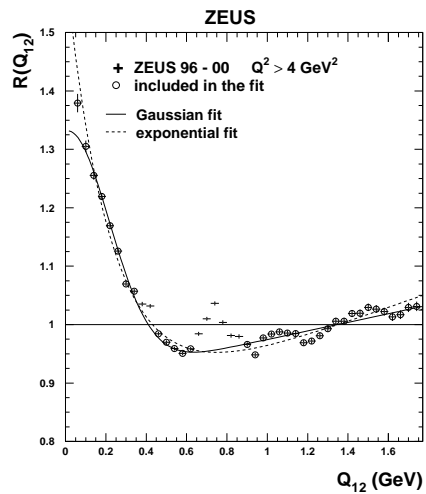}
\vspace*{-0.75cm}
\caption{\label{bose}Bose-Einstein correlations: 
ZEUS measurement of $R(Q_{12})$ for $Q^2 >$~4~${\rm GeV^2}$ 
together with two fits. The data points included in the
  fit are marked with circles.}
\end{center}
\end{figure}

\subsection{Bose-Einstein correlations}

ZEUS has measured Bose-Einstein (BE) correlations in
NC DIS over a wide kinematic range in $Q^2$ from 0.1~${\rm GeV^2}$ to
8000~${\rm GeV^2}$~\cite{zeusbe} in over 120~${\rm pb^{-1}}$ of data collected
from 1996 to 2000. 
As a measure for the correlation, the analysis used the double ratio 
$R(Q_{12}) = \zeta^{data}/\zeta^{MC,noBE}$, where $Q_{12}$ is the
Lorentz-invariant 
momentum difference of the particles (which are assumed to be pions), and
$\zeta^{data}$ ($\zeta^{MC,noBE}$) are the ratios of inclusive two-particle
densities for like-sign and unlike-sign bosons, $\zeta =
\rho(++,--)/\rho(+-)$, for data and Monte Carlo, respectively. The Monte Carlo
sample was generated without simulating Bose-Einstein effects, assuming
that BE correlations can be factorised from other types of correlations and
that non-BE correlations are well described by the MC models. Therefore 
$R(Q_{12})$ should be sensitive to BC correlations only.

Figure~\ref{bose} shows the double ratio $R(Q_{12})$ together with two fit
models which aim at extracting the size $r$ of the source of final state pions
and the coherence parameter $\lambda$. 
The data points marked with circles were actually included in the fit;
as expected the parameters extracted from the gaussian fit show no dependence
on $Q^2$. The radius of the production volume is found to be about 
$r \sim$~0.65~fm, and the coherence parameter $\lambda \sim$~0.47.
A two-dimensional study of the Bose-Einstein effect shows that the
pion-emitting region is elongated. All findings are compatible with
measurements from other experiments and colliders. 

\subsection{Anti-Deuteron production}

The H1 collaboration has measured anti-deuteron production~\cite{h1deuteron}
and compared the production rate to results from central proton-proton (Au-Au)
collisions measured at the CERN ISR (RHIC).  

No significant signal can be found for positively charged tracks,
but for negatively charged particles a clear
signal of 45 anti-deuterons with an estimated background of about 1 event is
observed, leading to a total anti-deuteron cross-section of about 2.7~nb. 

The production of nuclei in particle collisions can be described in terms of
the coalescence model in which the cross-section $\sigma_A$ for the production
of a nucleus with $A$ nucleons can be related to the cross-section for the
production of free nucleons in the same reaction, $\sigma_N$, by a so-called
coalescence parameter $B_A$. 

The (anti-)deuteron coalescence parameter value $B_2 \sim$~0.01 found by ZEUS
is compatible with values deduced at lower center-of-mass energies in $pp$ and
$pA$ collisions, 
but much larger than observed in $Au-Au$ collisions at similar nucleon-nucleon
center-of-mass energies. This is understandable if the size of the hard
reaction `fireball' is much smaller in $pp$ / $\gamma p$
than in heavy ion collisions.

\section{\label{conclusion}CONCLUSION}

Jet and hadronic final state physics at HERA has been reviewed. Clear progress
especially in jet physics in terms of statistical precision has been made over
the past few years. Jets have even been used in global QCD fits with the aim
of improving the knowledge of the gluon density at high values of $x$. 
In addition, exotic signatures like instantons or
pentaquarks have been analysed. Nevertheless, HERA physicists are eagerly
waiting for HERA 2 data which promise enough statistics to solve remaining
questions and to come to firmer conclusions concerning $\alpha_S$ and the
PDFs. 

\section*{ACKNOWLEDGEMENT}

I would like to thank my colleagues from H1 and ZEUS for their
support and the conference organisers for a pleasant stay in Montpellier.


\begin{thebibliography}{29}
\bibitem{refshape} ZEUS Coll., Abstract 5-0290, ICHEP 2004.
\bibitem{zeusprompt} ZEUS Coll., DESY-04-016.
\bibitem{zeusshapes} ZEUS Coll., DESY-04-072.
\bibitem{zeusglobal} ZEUS Coll., Abstract 5-0263, ICHEP 2004.
\bibitem{transition1} H1 Coll., DESY-03-160.
\bibitem{transition2} ZEUS Coll., DESY-04-053.
\bibitem{transition3} H1 Coll., DESY-03-206.
\bibitem{forwardjetsold} H1 Coll., Nucl.~Phys.~B538 (1999) 3;\\
  ZEUS Coll., Phys.~Lett.~B474 (2000) 1,233;\\
  ZEUS Coll., Eur.~Phys.~J.~C6 (1999) 239.
\bibitem{h1forwardnew1} H1 Coll., H1prelim-04-033.
\bibitem{h1forwardnew2} H1 Coll., DESY-04-51.
\bibitem{pqzeus} ZEUS Coll., DESY-04-164.
\bibitem{pqh1} K.~Lipka, these proceedings.
\bibitem{h1instanton} H1 Coll., Eur.~Phys.~J.~C25 (2002) 495.
\bibitem{zeusinstanton} ZEUS Coll., DESY-03-201.
\bibitem{zeusbe} ZEUS Coll., DESY-03-176.
\bibitem{h1deuteron} H1 Coll., DESY-04-032.
\end{thebibliography}
\end{document}